*Review:* **Tackling drug resistant infection outbreaks of global pandemic** *Escherichia coli* **ST131 using evolutionary and epidemiological genomics**

Tim Downing, School of Biotechnology, Faculty of Science and Health, Dublin City University, Dublin 9, Ireland.
Email: tim.downing@dcu.ie

**Abstract:** High-throughput molecular screening is required to investigate the origin and diffusion of antimicrobial resistance in pathogen outbreaks. The most frequent cause of human infection is *Escherichia coli*, which is dominated by sequence type 131 (ST131), a set of rapidly radiating pandemic clones. The highly infectious clades of ST131 originated firstly by a mutation enhancing virulence and adhesion. Secondly, single-nucleotide polymorphisms occurred enabling fluoroquinolone-resistance, which is near-fixed in all ST131. Thirdly, broader resistance through beta-lactamases has been gained and lost frequently, symptomatic of conflicting environmental selective effects. This flexible approach to gene exchange is worrying and supports the proposition that ST131 will develop an even wider range of plasmid and chromosomal elements promoting antimicrobial resistance. To stymie ST131, deep genome sequencing is required to understand the origin, evolution and spread of antimicrobial resistance genes. Phylogenetic methods that decipher past events can predict future patterns of virulence and transmission based on genetic signatures of adaptation and gene exchange. Both the effect of partial antimicrobial exposure and cell dormancy caused by variation in gene expression may accelerate the development of resistance. High-throughput sequencing can decode measurable evolution of cell populations within patients associated with systems-wide changes in gene expression during treatments. A multi-faceted approach can enhance assessment of antimicrobial resistance in *E. coli* ST131 by examining transmission dynamics between hosts to achieve a goal of pre-empting resistance before it emerges by optimising antimicrobial treatment protocols.

## 1. The advent of genome-based *Escherichia coli* monitoring

The rate of infectious disease outbreaks globally increased significantly during 1980-2013 [1,2]. Understanding the transmission dynamics of antimicrobial resistance (AMR) during infections necessitates deep molecular screening [3,4]. The most informative tools for epidemiological investigation of AMR exploit emerging high-throughput technologies [5], of which genome sequencing is becoming standard for resolving the origins of outbreaks over timespans ranging from centuries to days [6]. Only such tools can trace the transmission of infection outbreaks: other genetic profiling methods such as multi-locus sequencing typing (MLST) of housekeeping genes or pulse-field gel electrophoresis (PFGE) are not sufficiently discriminatory [7].

Although PFGE is still used internationally for reconstruction the evolution of bacterial outbreaks, its reproducibility between labs can be inconsistent [8] due to variability in genome-wide restriction enzyme site frequency. This can be reduced with use of standardised protocols, materials, and interpretive criteria. However, its low discriminatory power for ST131 [9] suggests other methods are superior [10].

MLST has moderate power to determine sequence type (ST) given sufficient elapsed time [11] but is limited by selecting a small number of genes from the genome [12] (Table 1). The Pasteur MLST system genes numbers eight genes: DNA polymerase (*dinB*), isocitrate dehydrogenase (*icdA*), p-aminobenzoate synthase (*pabB*), polymerase PolII (*polB*), proline permease (*putP*), tryptophan synthase subunits A and B (*trpA*/*trpB*) and beta-glucuronidase (*uidA*) (www.pasteur.fr/recherche/genopole/PF8/mlst/EColi.html). This differs from the Achtman scheme, which uses seven housekeeping genes: adenylate kinase (*adk*), fumarate hydratase (*fumC*), DNA gyrase subunit B (*gyrB*), *icd*, malate dehydrogenase (*mdh*), adenylosuccinate dehydrogenase (*purA*), ATP/GTP binding motif (*recA*) [13]. Over shorter timescales, MLST is limited [14]: genome sequencing can be considered simply as an extension of MLST that offers an expanded set of genes and thus finer resolution, which is critical for fighting rapid clonal expansions.

The comprehensive profiling provided by genome sequencing has enabled the production of new antimicrobials, diagnostics and surveillance systems, and is a fundamental shift in practice in infection control that enables more accurate pathogen tracing [15]. An example is the enteroaggregative ESBL *E. coli* O104:H4 summer 2011 outbreak that infected thousands and killed many in France and Germany [16]. Its source was initially misidentified (but later clarified as fenugreek seeds), and the incubation time in patients was eight days, and progression from diarrhoea to haemolytic-uremic syndrome a mere five days [17]. This indicates a short timespan for the medical application of molecular technologies. The draft genome was sequenced within three days and a consensus genome sequence was reconstructed within two further days [18], allowing validation of discovered AMR markers within 16 days [19]. In other examples, genome sequencing successfully determined the number of origins of ST131 CMT-X-15-positive infections in a neonatal unit [20] and in enterohaemorrhagic O157:H7 isolates [21]. Similar strategies have worked in clinical settings for other bacterial species like *Staphylococcus aureus* and *Clostridium difficile* [22].

Recent methodological advances have dramatically improved the accuracy and cost efficiency of high-throughput sequencing technologies, primarily in relation to genome sequence contiguity through better library preparation or genome assembly. At present, genomes can be amplified from just 20+ pg DNA [23] or even single colonies [24], and up to 384 samples can be sequenced simultaneously [25]. Sequencing single bacterial cells would not require prior cultivation, and therefore could detect cells on environmental surfaces [26], as well as removing biases associated with growing large numbers of cells, which favours those best-suited to media or that grow faster. Sequence depth and repetitive sequence biases can be reduced by using microwell displacement amplification systems. These have produced assemblies of over 90% of an *E. coli* genome [27]. *De novo* assembly of *E. coli* and other bacterial genomes as a single unit using long sequence reads is now viable using single-molecule real-time sequencing [28,29]. This approach can differentiate carbapenemase-producing elements in *E. coli* [30], determine the precise origins of the linked mobile genetic elements (MGEs) like insertion sequences (ISs) [31], and also create gapless enterohaemorrhagic O157:H7 EDL933 genome with no ambiguous SNPs [32]. Additionally, other strategies like nanopore sequencing have produced high-resolution *E. coli* genomes [33].

## 2. *Escherichia coli* ST131 is a major global health issue

*E. coli* is the most frequent cause of acute bacterial infections [4], particularly blood stream infections (BSIs) and urinary tract infections (UTIs) [34]. *E. coli* causing UTIs and BSIs are not genetically distinct and can transfer between distinct environments [35]. Uropathogenic *E. coli* cause 80% of UTIs [36] and infect 20% of adult women at least once in their lifetime [37]. Approximately 50% of hospital-acquired (HA) and 70-95% of community-acquired (CA) UTIs are caused by *E. coli,* and these are suspected to originate from intestinal colonisers [37,38]. There are multiple potential sources of *E. coli* UTIs and BSIs because it contaminates food [39,40,41] such as fruit [42] and meat [43], and infects companion animals [44] like cats [45] and dogs [46].

*E. coli* are the most frequent extended spectrum beta-lactamase (ESBL)-producing bacterial species and are common globally: ESBL-producers account for 19-61% of *E. coli* in Spain [47], 41% in Israel [48], 41% in Japan [49], 47% [50] to 56% [51] in the USA, 78% in Canada [52] and 91% in Ireland [53]. The most significant threat among ESBL *E. coli* is sequence type O25b:H4-B2-ST131, a set of pandemic clones from phylogroup B2 that is dominant worldwide [54,55,56]. B2 is associated with extraintestinal infection, but is also present asymptomatically in humans [57,58]. There is evidence that B2 diverged early among *E. coli* groups and is composed of nine subclasses [59], of which ST131 is a basal group [60]. O25b ST131 has been detected in many countries, and serotypes O16:H5, O(not typeable):H4, and O157 have also been observed in Australia [61], China [62], Denmark [63], Japan [64], Spain [47], the USA [65] and the UK [54,66].

Even though it was first reported in 2002 [67], ST131 is not new. Global ST131 isolates taken from humans, animals and environmental locations evaluated using PFGE include samples from 1967, 1982, 1983, 1985, 1986, and 22 from 1990-99 [50]. The earliest other ST131 are from Sweden and Britain and date to 1968, with others from the 1970s (7), 1980s (9) and 1990s (16) [68] – as well as one from 1985 [65].

Bacterial AMR is a major threat to public health [69]: AMR in *E. coli* is evident worldwide [4] and encompasses many compounds [24,70]. AMR in *E. coli* is becoming more frequent [71] and is driven by environmental exposure [72], such as in effluent waste water [73,74] and soil treated with manure from antibiotic-treated livestock [75]. In the UK from 1991-2012, antibiotic treatment failure jumped by ~12% and was more frequent in certain second-line antimicrobials (such as quinolones and cephalosporins) than in first-line ones (penicillins, macrolides, flucloxacillins) [76]. 75% of *E. coli* from 1997-2007 in Irish hospitals displayed AMR to eight or more of 16 antimicrobials [77]. ESBL-producing *E. coli* are resistant to cephalosporins and fluoroquinolones, leaving carbapenems as the sole last-resort antimicrobial [4]. However, multi-AMR ESBL *E. coli* with carbapenem-resistance are now detected with alarming frequency [78] and in Algeria [79], China [80] and Ireland [81] for ST131.

ST131 represents a universal problem whose evolutionary epidemiology needs deeper study. Little is known about how it evolves and spreads on a local scale [82] despite a high infection rate in neonatal [20,83,84,85] and childcare facilities [86]. As a fraction of total *E. coli* infections, ST131 infection rates are proportionally higher in long-term care facilities (LTCFs) (76%) than hospitals (49%) and the public (15%) [87]. This may be driven by the acquisition of CTX-M (cefotaximase) elements, which are beta-lactamases (*bla*) that hydrolyse beta-lactam rings [88]: UTIs caused by *bla*CTX-M-positive ST131 are now common in LTCFs [89], hospitals [90] and the community [91]. Current hypotheses on the sources of *E. coli* have identified LTCFs as a closed microenvironment in which AMR bacteria evolve [92], and subsequently they diffuse into the wider community, spreading AMR alleles [93].

## 3. The genomic landscape of antimicrobial resistance in *E. coli*

Commonly *E. coli* genomes are separated into the largely chromosomal highly-conserved core genome and the accessory genomes. Notably, the core genome varies between studies [94] because it is composed of genes present in all samples for that analysis [95]. The accessory genome typically has much lower sequence conservation and encodes non-essential traits associated with virulence and antimicrobial metabolism [96]. It represents a major component of microbial variation and includes MGEs like plasmids, transposable elements (TEs), pathogenicity islands, and prophages.

*E. coli* undergo extensive horizontal gene transfer (HGT) [97]: HGT accounts for ~31% of genome-wide variation in ST131 [98]. Genes arising by HGT are frequently associated with duplications, which can permit new functions among the two gene copies [99]. HGT typically occurs through three processes: the primary one is transduction via bacteriophages after cell lysis and invasion via homologous recombination. ST131 acquire AMR through extrachromosomal MGEs from bacteriophages [100], or the integration of prophage DNA into chromosomes [101]. Prophage elements account for over 35% of *E. coli* CDS variation [102], and encode genes linked to virulence [103], growth during nutrient deficit [104], transcriptional regulation [105], AMR [106] and biofilms [107]. During generalised transduction, HGT segment length is a function of phage type. Other options for HGT are the transformation of local DNA from dead cells, or conjugation mediated by plasmids, TEs, integrons or integrating conjugative elements (ICEs). Homologous recombination between ICEs increase their diversity [108]. TEs such as ISs have a high frequency of integration and excision, and mediate both AMR gene composition and their expression rates. Although HGT is generally a function of sequence homology, but TEs do not need high levels to insert.

More detailed knowledge of AMR gene types and expression rates can help optimise antimicrobial treatments [109,110]. A large number of ST131 genomes have been published. The first was NA114 from an Indian UTI: a 4,936 Kb genome with a GC content of 51.2% and of which 88.4% was coding sequence (CDS) [111]. In comparison, genome JJ1886 from a UTI was 5,130 Kb and genetically more representative of the super-spreading subgroups within ST131 [112]. Notably, JJ1886 contained a chromosomal CTX-M-15 element due to a partial insertion of a Tn3 element into a lambda-like prophage locus. Isolate EC958 from the UK was also *bla*CTX-M-15-positive and provided a 50.7% GC 5,110 Kb genome somewhat distinct from NA114 and JJ1886 [113], perhaps due to its earlier isolation date. It is important to note that the number of genes, tRNAs and rRNAs in NA114 (4,875; 67; 3), JJ1886 (5,086; 88; 22) and EC958 (4,982; 89; 7) may differ due to differences in sequence library preparation and genome assembly: comparisons must be cognisant that not all genomes are created equal.

## 4. Antimicrobial resistance genetically defines ST131

The most closely related clonal complexes (ST1876 and ST95) provide a genetic definition for ST131 (Figure 1). ST131 possesses a variety of genes associated with virulence [47,114] and pathogenesis [115] (Table 1). A key one is the *fimH* gene encoding a type 1 fimbrial adhesion protein expressed on the cell membrane. ST131 encompasses firstly serotype O16:H5 with a *fimH41* allele: this is equivalent to Clade A from Petty et al [116]. ST131 secondly includes serotypes O25b:H4, some of which possess *fimH22* allele, corresponding to Clade B [116]. Genomes from 1967, 1983 (avian infection), 1985, 1990 (avian), 1992, 1995 (chicken) and 1997 were all fluoroquinolone-sensitive *fimH22* (Clade B) ST131 [98]. Genome sequencing and the Pasteur MLST system can distinguish O16:H5 from O25b:H4, but not the Achtman MLST scheme.

The remaining ST131 belong to a group named H30 (or Clade C [116]) defined by their acquisition of a *fimH30* allele: the earliest date of appearance of H30 is at least 1997 [50]. A minority of H30 appear to have lost *fimH30* for *fimH22/35* or rarer *fimH* variants [118]. Fluoroquinolone-sensitive ST131 was common prior to 2000, but is now rare. This is due mutations in two genes encoding DNA gyrase subunit A (*gyrA*) and DNA topoisomerase 4 subunit A (*parC*). Nested within H30 is H30-R, which accounts for 58% of ST131 and is becoming more prevalent [116]. H30-R is nearly always fluoroquinolone-resistant – but this can be lost [118].

A subgroup of H30-R named H30-Rx is a recent radiation dating to at least 2002 and is nearly always *bla*CTX-M-positive [98,116]. Just three genome-wide SNPs distinguish H30-Rx from other H30-R. It is the main driver of increasing rates of fluoroquinolone-resistance [120]. *Bla*CTX-M-15 is the most frequent H30-Rx CTX-M element type and originated from *Kluyvera* in the early 2000s [119], such that 49% of *E. coli* LTCF samples from 2004-2006 were *bla*CTX-M-15-positive ST131 [121]. Other older fluoroquinolone-resistant *bla*CTX-M-15-positive H30-Rx with published genomes predating 2006 are EC958 (March 2005) [113], several from 2000-05 isolated in Canada [116], and from 2002 again in Canada, but also 2003 in Korea and 2005 in Portugal [98]. Such isolates could be essential for determining the origins of H30-Rx. *Bla*CTX-M-15 is also found less commonly in ST38, ST405 and ST648 [120].

Figure 1. ST131 genetic groups and acquisition of drug-resistance changes.

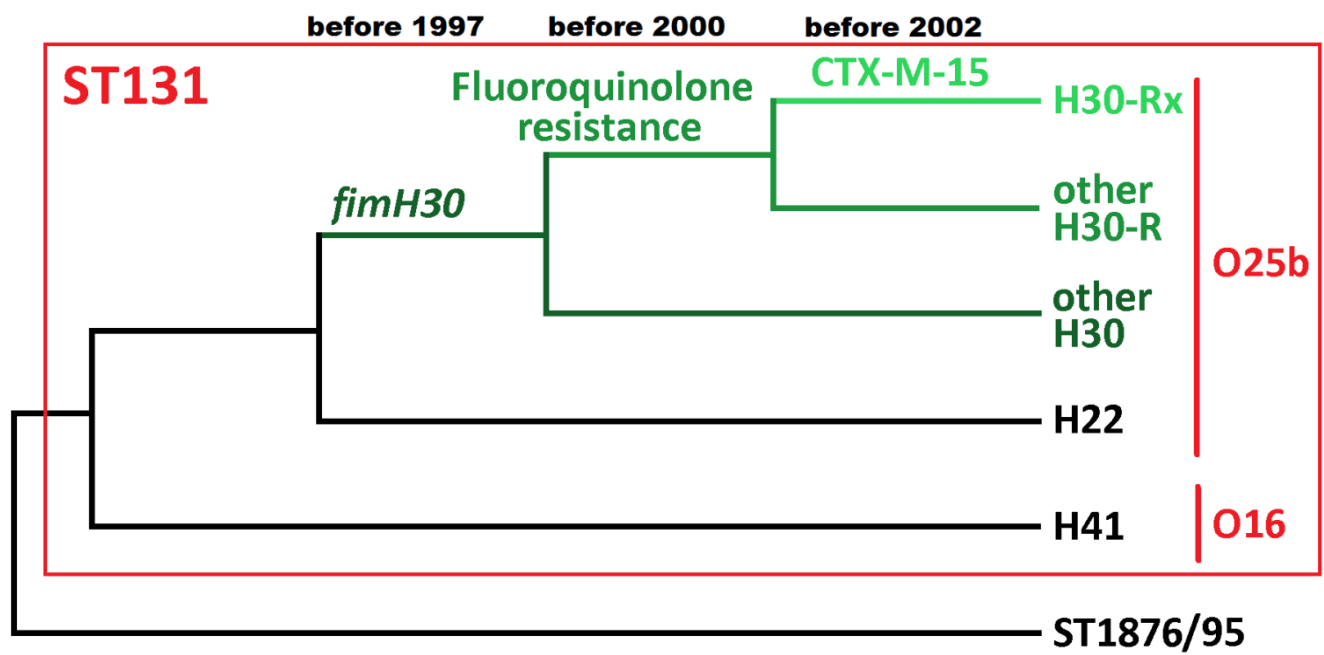

This cladogram shows ST1876 and ST95 as the most closely related clonal complexes to ST131 (with published genomes). H41 is in serotype O16:H5 and is known as Clade A (H41) [116]. H22 is a set of O25b:H4 ST131 called Clade B [116]. H30 is a subgroup of O25b:H4 ST131 defined by their acquisition of the *fimH30* allele (dark green) [98]: this is called Clade C. H30-R is a subset of H30 that are nearly always fluoroquinolone-resistant (emerald green), largely due to alleles 1AB in *gyrA* and 1aAB in *parC*. Within this group, H30-Rx have *bla*CTX-M-15 (bright green) [98] but not always [116]. This diagram does not mean that H30, H30-R or H30-Rx are monophyletic [68].

## 5. The key antimicrobial resistance elements in ST131

Numerous core genome mutations are implicated in ST131's AMR [122]. These include regulators of drug efflux, repressor of the *marRAB* (multiple antibiotic resistance) operon (*marR*), and repressor of *acrAB* (achromobactin outer membrane receptor) genes (*acrR*), that respond to drugs associated with high fitness costs. Others are L42R in ribosomal gene S12 (*rpsL105*); a transposon mediating resistance to tetracycline (*Tn10*); S80I and E84V in *parC* to fluoroquinolones; ciprofloxacin and nalidixate thanks to changes at *gyrA* (most frequently S83L and D87N) [123]. Chromosomal *ampC*-like cephalosporinases are also common in *E. coli* [48].

ST131's virulence genes are used to categorise it into distinct virotypes [68]. The aerobactin receptor (*iutA*), group 2/3 capsule synthesis (*kpsM II/III*) and *fimH* genes are part of a system for defining extraintestinal *E. coli* [124,125]. These are markers for H30-Rx, along with Afa adhesion (*afa*), Dr-binding adhesion (*dra*), virotype A and capsule (K100) genes [68]. ST131 virulence factor genes are typically located in pathogenic islands or MGEs and encode toxins, adhesions, lipopolysaccharides, polysaccharide capsules, proteases, and invasins (Table 1). Other genes universal in ST131 include secreted autotransporter toxin (*sat*), yersiniabactin receptor (*fyuA*), uropathogen-specific protein (*usp*), pathogenicity island marker (*malX*), adhesion receptor (*iha*), outer membrane receptor (*ompT*), aerobactin (*iucD*) and serum resistance associated (*traT*) [126].

ST131 has different plasmids from multiple incompatibility groups (Inc) such as IncF, IncI1, IncN and IncA/C. IncF are limited to *Enterobacteriaceae*, are the most frequent plasmids in ST131 [127], and have a major role in determining AMR. Typically they are >100kb long and maintained at a low number of copies. Three types of IncF plasmids have been found in ST131, not including complex fusions: IncFIA, IncFII and IncI1 [127]. These plasmids typically possess multiple AMR, beta-lactamase, virulence, toxin and antitoxin elements. Conjugation of IncFII plasmids have facilitated the pervasive spread of *bla*CTX-M elements, especially *bla*CTX-M-15. A majority of isolated ST131 have IncFII pEK516 plasmids, which has 103 genes including *bla*CTX-M-15, *bla*OXA-1, *bla*TEM-1, chloramphenicol acetyltransferase (*catB4*), tetracycline efflux pump (*tetA*) and plasmid stability (*stbB*) for stable plasmid inheritance [128]. In the latter study, plasmid pEK516 also contained aminoglycoside acetyltransferases allowing resistance to aminoglycosides, ciprofloxacin (both via *aac(6')-Ib-cr*) and gentamicin (via *aac(3)-II*).

Plasmids contain genes encoding various beta-lactamases that are grouped into Ambler classes A (*bla*CTX-M, *bla*TEM, *bla*KPC, *bla*SHV); B (*bla*VIM, *bla*NDM, *bla*IMP); C (*bla*CMY, *ampC*-related like *bla*DHA); D (*bla*OXA). Classes A/C/D use serine for beta-lactam hydrolysis, whereas B are metallo-beta-lactamases that use divalent zinc ions [129]. Beyond *bla*CTX-M in class A, TEM beta-lactamases are the most frequent ESBL in *E. coli* more broadly. They are exchanged extensively between species: for example, TEM-52C in *E. coli* is mediated by an IncI1-g plasmid on a Tn3 transposon [130] and has been found in *Salmonella* [131]. SHV beta-lactamases are less common in *E. coli*, but are abundant in *K. pneumoniae*, from which ST131 has acquired other beta-lactamases [132].

Class A KPC-2 (*K. pneumoniae* carbapenemase) is the most common carbapenemase type [120]. This can impair cephalosporins, monobactams and penicillins, as well as carbapenems, and has spread to China [80], France [133], Ireland [78,81] and the USA [134]. Other carbapenemases include class B metallo-beta-lactamases like Verona integron-encoded metallo-beta-lactamase (VIM), which has been found in Italy [135], and *bla*IMP-8, which has been detected in Taiwan [136].

Class B NDM-1 (New Delhi metallo-beta-lactamase)-positive ST131 has been found in India [137] where it may have originated in *E. coli* by *in vivo* conjugation with *K. pneumoniae* in 2009 or earlier [132]. *Bla*NDM-1-positive ST131 is now common in northern India (6% in Varanasi [138]), southern India (7% in Mumbai [139]) and northern Pakistan (15% [140]). *Bla*NDM-1 has been found on fast-spreading IncFII plasmids as well, along with Class D carbapenemase *bla*OXA-1 (oxacillinases) together with aminoglycoside resistance genes [141]. The rapid spread of *bla*NDM-1 in many species provides an example of how easily certain AMR genes can be spread.

Lastly, class C includes plasmid-mediated *ampC* beta-lactamases *bla*DHA-1 [49], which was first found in *Salmonella enterica* [142], and *bla*CMY-4 (active on cephaycins) [48] and *bla*CMY-2 [49,143] – the latter is a cephamycinase originally transferred from *Citrobacter* species. Some *ampC* genes are chromosomal, but possess a much lower expression rate than the more common plasmid-mediated *ampC* genes [48].

## 6. How did ST131 adapt to be so successful?

Bacterial infectious disease is driven in part by rapid mutation rates (~$10^{-5}$ per genome per generation in *E. coli*) [144], short generation times [145], and non-lethal antimicrobial doses due to non-adherence [146]. Remarkably, AMR may have no fitness cost and may even increase fitness in *E. coli* [147,148]. ST131 is highly resistant without any fitness cost [127], even more so in CA compared to HA ST131 isolates [149]. As outlined, the three key changes define H30-Rx ST131: the acquisition of *fimH30*, fluoroquinolone-resistance switches and beta-lactamase gains. In contrast to the single acquisitions of *fimH30* and fluoroquinolone-resistance, the type of beta-lactamase varies, and they are lost and recovered frequently. Consequently, antimicrobial treatments have inadvertently created the H30-Rx superbug, and associated fitness costs may be linked with beta-lactamases, but less so for *fimH30*, 1AB in *gyrA* and 1aAB in *parC*. As acknowledged above, there are rare cases of *fimH30* loss and *gyrA/parC* mutation within H30-Rx.

**(i) Extensive horizontal gene transfer**
ST131 is common in UTIs and BSIs, but not at other anatomical locations [150]. In spite of pervasive HGT, DNA exchange with other bacterial species is rare, affecting just ~0.4% of the core genome conserved across the *E. coli* genus [151]. UTI-causing ESBL-positive ST131 acquire resistance through HGT with other ST131 colonising the same individual, but rarely from other *E. coli* phylogroups [54]. Indeed, ST131 is frequent in poultry [152] but retains a distinct set of *bla*CTX-M and *bla*TEM-52 elements despite high overall genetic similarity [153]. Naively, this suggests previous HGT with *Kluyvera* [119] and *K. pneumoniae* [132] is rare. However, recent examples of *K. pneumonia* derived plasmids contributing *bla*CTX-M-27 genes to ST131 [154] is consistent with a proposal of alarming potential genetic flexibility in ST131.

**(ii) Maintenance of a broad resistome**
ST131 regulates the activity of its resistome: its entire complement of AMR genes. There is evidence of functional robustness and redundancy in the resistance mechanisms of *E. coli*'s "proto-resistome" comprising penicillin-binding proteins, cell wall modifying enzymes and cell division genes [155]. A sequencing approach could assess this in a comprehensive manner. Expression of 61 genes is associated with small increases in tolerance to 86 antimicrobial-related compounds [156]. Many potential AMR genes remain unknown in *E. coli* [157], even though their activities vary even within putatively clonal cell populations [158] and the mechanism of action of many antimicrobials remains unclear [159].

**(iii) Regulatory fine-tuning of gene expression**
Regulatory sequences alter AMR through gene expression levels [160] including bursting [161] and transitions between pathogenic and non-pathogenic states [126]. Promoters evolved to alter gene expression rates in response to rapidly changing environmental conditions [162], which can affect AMR phenotypes by up to $10^6$-fold [163]. Promoter gene expression regulation is higher at non-essential genes with lower sequence conservation, and their activity rates are more variable than those of essential genes [164]. During multi-drug exposure, 30% of *E. coli* expression variation is attributed to promoter mutations [165] – for example, a mutation T32A at a beta-lactamase *ampC* gene promoter elevates expression rates genome-wide [166].

**(iv) Cross-antimicrobial resistance and compensatory mutations**
Synergy and antagonism between different antimicrobials is pervasive [167], even to the point that simultaneous synergy and antagonism is possible (a Harvey Effect). Some compensatory mutations are specific to individual antimicrobials but others promote cross-AMR simultaneously [168]. Feedback-based cross-AMR in response to combination treatments is a function of the number of unlinked (positive) gene regulatory networks [169]. AMR is reduced by negative epistasis during cyclical treatment, which decreases resistance to both drugs more effectively than to single drugs – nearly as well as dual drug therapies [170]. Certain compensatory mutations alter AMR without being clearly associated with the phenotype – for instance, *bla*CTX-M-linked carbapenem-resistance requires water channel (porin) gene loss to mitigate fitness costs [171]. Additionally, structural rearrangements at ISs are associated with rapid increases in fitness during long-term *E. coli* evolution [172].

**(v) Resistance through cell growth arrest**
Antimicrobials limit *E. coli* growth [173], which is controlled by gene expression [174]. Growth depends on environmental conditions and is controlled by the concentration of transcription-associated proteins [175]. Even without a specific stress, cells spontaneously enter and exit a quiescent non-growth state during both log and stationary growth phases [176]. Lowered expression is more strongly associated with non-essential genes like toxins [164]. There are 22 toxin-antitoxin systems in *E. coli* K12 that can be recruited during environmental stress to regulate growth, affecting biofilm formation [177]. For example, in the *hipBA* toxin-antitoxin system, antitoxin HipB represses serine-threonine-protein kinase HipA but has a short half-life. When HipB levels fall, HipA phosphorylates a glutamyl tRNA synthetase to mimick starvation [178], reducing translation and slowing growth, resulting in a non-growing phenotype [179].

**7. Stopping ST131:**

**(i) Evolutionary phylogenomics and fitness measurement**
Pathogens resist antimicrobials by evolving: consequently, analysis of evolution is needed [180] which should operate on a genome-wide basis for outbreaks like ST131 to more finely resolve of evolutionary patterns at all potential AMR factors [181]. Genome-based evolutionary analysis and real-time diagnostic evaluation is now as cheap and time-effective as conventional approaches [182] and informative bioinformatics strategies now underpin improvements in clinical microbiology [183].

Variation arising by recombination between strains can be deduced from the genome-wide DNA mutation density distribution [181]: signatures of HGT have much higher polymorphism rates, reflecting the mutational time spent in a different species [184]. For H30-Rx, this includes MGEs, integrated phage DNA and capsule-related genes [185]. Comparing the most likely genealogy at a given panel of SNPs (haplotype) to the mean signal across the genome can identify ancestral and recent recombination between and within subpopulations [186] and determine the phylogeographic origin of specific recombination events in a coalescent framework [187]. Phylogenetic and genealogical histories at regions arising through HGT deviate substantially from the mean genome-wide phylogenetic signal [188,189].

New beneficial variants increase in frequency faster than average [190] and retain a more recent time of origin [191]. Phylogenetic branching topology can measure relative fitness assuming a persistent source of antimicrobial pressure causing mutations with small fitness effects [192]. The latter is valid for *E. coli* because many genes regulate AMR and virulence [156,193]. No variation in relative fitness would mean low variability in the rates of coalescence of descendent nodes compared with ancestral ones. In contrast, a high difference in relative fitness would mean highly fit ancestors will produce many more descendent nodes. This would appear as a radiating population: such as H30 within ST131, H30-R within H30, H30-Rx with H30-R; what will follow H30-Rx?

**(ii) Modelling historical and future epidemiology**
Systematic monitoring of infection and transmission in human populations can be used to infer sources of ST131, including from potential reservoirs like livestock and intestinal colonisers [61]. Current hypotheses pinpoint the Indian subcontinent as a possible original reservoir for ST131. During the discovery of ST131 in 2008, there was a high rate of recent travel to this region in those with H30-Rx ST131 infection in New Zealand [194] and Canada [195,196]. Whether this was caused by both a single H30-Rx clone and through plasmid-mediated HGT is unclear [197] due to the limited sampling of the Indian subcontinent prior to 2002. With extensive geographic sampling including samples particular during the acquisition of *fimH30* in the 1990s and fluoroquinolone-resistance sometime before 2002, phylogenomic reconstruction would address this puzzle.

Genome sequencing can help address the rise of AMR [38] by predicting future resistance. The most likely future dominant subtype of a pathogen is typically the most fit one [198]: this can be applied to ST131 based on *de novo* and known mutations at key AMR determinants using frameworks developed for viruses [199]. The relative rate of allele frequency change reflects its association with antimicrobial selection pressure [200], and can be measured from time-series genome data [201] to distinguish this effect from drift [202]. Such sampling over extended periods allows the identification of genes implicated in adaptation based on the mutation's age, functional effects, and ancestral phylogenetic position [203].

Phylodynamic models determine future evolutionary and epidemiological patterns based on phylogenetic tree structure [204]. Such schemes have revealed missed events during transmission of tuberculosis [205], pin-pointed an asymptomatic spreader of methicillin-resistant *S. aureus* between babies [206], assessed the potential for multiple simultaneous outbreaks to originate from the same common ancestor [207], and examined evidence for multiple origins of infection [208]. Although such methods have not yet been applied to *E. coli*, they can enhance our understanding of the inter-host spread of AMR.

### (iii) Assessing cell growth arrest
Stochastic phenotype switching is common to all domains of life, is accentuated by antimicrobial stress, and in bacteria is observed as changes in gene expression. This growth-dormancy bistability is an evolutionary bet-hedging strategy that results in a mix of cells with varying gene expression rates [209]. It arises due to gene regulatory network structure, leading to rare but occasional switches from the growing wild-type (WT) state to a non-growing dormant one. Cell-cell interactions may affect variation in gene expression rates [210], allowing this randomness to adaptively drive regulatory changes.

Cell dormancy causes large shifts in metabolic gene expression [211,212]. This contrasts with ribosomal gene expression that has buffering mechanisms [174,213] and so scales linearly with cell growth [214]. *E. coli* alter their carbon metabolism rapidly [215] by co-regulating enzymes to cope with rapidly changing environmental conditions [216]. However, metabolic flux is regulated extensively [217,218] so differing concentrations can produce the same net metabolic effects [219,220] or affect other pathway components [221]. *E. coli* optimise gene expression by moderating ribosome production [222]: this means shifts in metabolic gene expression can be compared to those for ribosomal RNA, which should match the antimicrobial toxicity and bacterial AMR level. AMR-driven changes in gene expression vary non-monotonically with dose and are a function of the complexity of the mechanism: antimicrobials inactivating targets may elicit higher target expression, whereas drugs causing gain-of-function changes may cause mutation [110]. This means the relative rates of metabolic and ribosomal gene activity provide sufficient information to deduce AMR levels and types.

### (iv) Avoiding partial treatments and cell population heterogeneity
Suboptimal treatment regimens [34], poor compliance [223] and drug pharmacokinetics [224] lead to spatial structure in antimicrobial concentrations. This is frequent in human and animals both at an individual level [225] and in groups of treated and untreated patients [226]. At low antimicrobial doses (such as a minimal selective concentration [227]), WT and resistant cells grow equally well, whereas only resistant cells would survive when exposed to dose greater than the minimal inhibitory concentration (MIC). Partial treatment regimens accelerate AMR for mechanisms requiring numerous mutational steps with small fitness costs by offering a sub-MIC sanctuary where resistant cells can evolve further despite transiently lowered fitness [226]. However, partial regimens can slow AMR if it requires few mutational steps or substantial temporary falls in fitness [229]. In addition, the mutational time required for resistance emergence may be a function of cell density [230].

Spatial or temporal variation in treatment application allows WT cells more time to accrue new mutations, causing faster AMR evolution [228]. These genetically resistant cells are known as type 1 persisters to differentiate them from type 2 persisters that are resistant due to gene expression control causing non-growth [231]. Although *E. coli* type 2 persister gene activity depends on adjacent cells [165], they revert to growth after dormancy ends, and attain exponential growth rates once resistant mutations occur when all WT cells have died [232]. These type 2 persisters may require a genetic predisposition [231] and initially comprise just 0.001% of cells [233].

**(v) Dissecting measurably evolving infections**
A single ST131 infection may become a genetically diverse cell community descended from a recent common ancestor driven by antimicrobial selective pressure. Multiple populations could exist: for example *S. aureus* measurably evolves (30 SNPs and 4 indels in 16 months) to magnify virulence [234]. Furthermore, distinct *Burkholderia dolosa* lineages co-exist within individual cystic fibrosis patients for years and have undergone extensive clonal interference during exposure to antimicrobials [235]. Soft selective sweeps during host adaptation of *Pseudomonas aeruginosa* are prevalent, indicating different loss-of-function mutations may target the same gene within a heterogeneous cell community [236].

Consequently, deep genome sequencing is required to grapple with mutations in a cell population [237]. Such population mixes could be described using birth-death [181], logistic growth [238] and two-level population models within a single infection [239] or a structured environment [240] to optimise antibiotic treatment protocols [241]. However, there are caveats: high-dose antimicrobial treatment may give rise to a single dominant variant within the host [110], rare variants may become more advantageous at differing rates as the host microenvironment changes [190], and mutants may be pre-existing rather than *de novo* [123].

**Conclusion: Future avenues for ST131 infection genomics**

Genome sequencing can predict virulence, toxicity and AMR phenotypes in *E. coli* ST131 [242]. It is a pivotal tool for infection control because it facilitates the decoding of molecular mechanisms of treatment resistance [243] and can dissect clonal outbreaks of monomorphic bacteria [244]. It can decipher transmission routes and inter-host contact [181], and evaluate genetic diversity within hosts over time [245]. This evolutionary approach should become the basis for analysing *E. coli* ST131 outbreaks. A final point is that there are several other possibilities for genome-based AMR control not discussed here: *E. coli* post-transcriptional processing disturbs expected correlations between mRNA and protein levels [246] – this includes AMR-driven regulatory changes by small RNAs [247]. Additionally, epigenetic heterogeneity within *E. coli* cell populations screened using long-read sequencing methylation data [248] is associated with extensive regulatory control of phenotypes [249].

**Acknowledgments**

I thank Cathal Seoighe, Martin Cormican, Dearbhàile Morris and Catherine Ludden (all NUI Galway) for discussions. This work was supported by the National University of Ireland Galway (Ireland) and USA National Science Foundation (Grant No. NSF PHY11-25915).

**Author Contributions**
T.D. conceived and wrote the review.

**Conflicts of Interest**
The author declares no conflict of interest.

Table 1. Genes associated with virulence or ST131 typing. Genes encoding virulence factors (adhesion, capsules, siderophores, antimicrobial resistance (AMR), toxins) and used to define Extraintestinal *E. coli*, H30-Rx ST131 and the Pasteur/Achtman MLST schemes are highlighted.

| Gene Name | Gene product | Role in virulence | Typing use |
|---|---|---|---|
| H7 fliC | Flagellin variant | Multiple | |
| iss | Increased serum survival | Multiple | |
| malX | Pathogenicity island marker | Multiple | |
| ompT | Outer membrane protease T | Multiple | |
| traT | Serum resistance-associated | Multiple | |
| usp | Uropathogenic-specific protein | Multiple | |
| K1/K2/K5 | Group 2 capsule variants | Capsule-related | |
| kpsM II | kpsM II group 2 capsule | Capsule-related | Extraintestinal *E. coli* definition |
| kpsMT III | Group 3 capsule | Capsule-related | Extraintestinal *E. coli* definition |
| K100 | Group 2 capsule variants | Capsule-related | H30-Rx definition |
| afa | Afa adhesin | Adhesion | H30-Rx definition |
| draBC | Dr-binding adhesin | Adhesion | H30-Rx definition |
| fimH | Type 1 fimbriae | Adhesion | Extraintestinal *E. coli* definition |
| F10 papA | P fimbriae subunit variant | Adhesion | |
| focG | F1C fimbriae adhesin | Adhesion | |
| hra | Heat-resistant agglutinin | Adhesion | |
| iha | Adhesion siderophore | Adhesion | |
| papC/papE/papF/papG | P fimbriae operon | Adhesion | |
| sfa/foc | F1C fimbriae or S | Adhesion | |
| tsh | Temperature sensitive hemagglutinin | Adhesion | |
| iutA | Aerobactin receptor | Siderophore | Extraintestinal *E. coli* definition |
| fyuA | Yersiniabactin receptor | Siderophore | |
| ireA | Siderophore receptor | Siderophore | |
| iroN | Salmochelin receptor | Siderophore | |
| stbB (pEK516 plasmid) | Plasmid stability | plasmid stability | |
| catB4 (pEK516 plasmid) | Chloramphenicol acetyltransferase | AMR | |
| tetA (pEK516 plasmid) | Tetracycline efflux pump | AMR (tetracycline) | |
| aac (pEK516 plasmid) | Aminoglycoside acetyltransferases | AMR (multiple) | |
| mphA | Macrolide 2'-phosphotransferase I | AMR (macrolide) | |
| acrR | Repressor of acrAB genes | AMR (multiple) | |
| marR | Repressor of the marRAB operon | AMR (multiple) | |
| rpsL105 | Ribosomal gene S12 | AMR (multiple) | |
| aadA5 | Aminoglycoside 3'-adenyltransferase | AMR (streptomycin) | |
| sulI | Dihydropteroate synthase | AMR (sulfonamide) | |
| Tn10 | Transposon - tetracycline resistance | AMR (tetracycline) | |
| dfrA7 | Dihydrofolate reductase type VII | AMR (trimethoprim) | |
| gyrA | DNA gyrase subunit A | AMR (fluoroquinolone) | |
| parC | DNA topoisomerase 4 subunit A | AMR (fluoroquinolone) | |
| astA | Arginine succinyltransferase | Toxin | |
| cnf1 | Cytotoxic necrotizing factor | Toxin | |
| hlyD | Alpha-Hemolysin | Toxin | |
| pic | Serine protease | Toxin | |
| sat | Secreted autotransporter toxin | Toxin | |
| vat | Vacuolating toxin | Toxin | |
| dinB | DNA polymerase | | Pasteur MLST system |
| pdbB | P-aminobenzoate synthase | | Pasteur MLST system |
| polB | Polymerase PolII | | Pasteur MLST system |
| putP | Proline permease | | Pasteur MLST system |
| trpA/trpB | Tryptophan synthase subunits A, B | | Pasteur MLST system |
| uidA | Beta-glucuronidase | | Pasteur MLST system |
| icd/icdA | Isocitrate dehydrogenase | | Pasteur/Achtman MLST system |
| adk | Adenylate kinase | | Achtman MLST system |
| fumC | Fumarate hydratase | | Achtman MLST system |
| gyrB | DNA gyrase subunit B | | Achtman MLST system |
| mdh | Malate dehydrogenase | | Achtman MLST system |
| purA | Adenylosuccinate dehydrogenase | | Achtman MLST system |
| recA | ATP/GTP binding motif | | Achtman MLST system |